\documentclass[twocolumn,showpacs,preprintnumbers,superscriptaddress]{revtex4-1}
\usepackage{amsmath}
\usepackage{amssymb}
\usepackage{amsmath}
\usepackage{epsfig}
\usepackage{graphicx}
\usepackage{inputenc}
\inputencoding{latin1}
\usepackage{ulem}

\ifx\hypersetup\sadfkjashdfkxja\else
\hypersetup{pdfsubject={}}
\hypersetup{pdfkeywords={}}
\hypersetup{plainpages=false}
\hypersetup{bookmarksnumbered=true}
\hypersetup{pdfstartview=FitH}
\hypersetup{pdfpagemode=UseNone}
\hypersetup{colorlinks=false}
\hypersetup{citebordercolor={.5 1 .5}}
\hypersetup{urlbordercolor={.5 1 1}}
\hypersetup{linkbordercolor={1 .7 .7}}
\fi

\newcommand{\ud}{\mathrm{d}}
\newcommand{\be}{\begin{equation}}
\newcommand{\ee}{\end{equation}}
\newcommand{\bea}{\begin{eqnarray}}
\newcommand{\eea}{\end{eqnarray}}

\newcommand{\Appendix}[1]%
    {%
     \section{#1}%
      }



\begin{document}
\title{A method for getting well-defined coupling constant in all region}
\author{Zhi-Yuan Zheng}
\email{zhengzy@itp.ac.cn}
\affiliation{Key Laboratory of Theoretical Physics, Institute of Theoretical Physics, Chinese Academy of Sciences\\ Beijing 100190, People's
Republic of China}
\affiliation{School of Physical Sciences, University of Chinese Academy of Sciences\\No.19A Yuquan Road, Beijing 100049, China}
\author{Gao-Liang Zhou}
\email{zhougl@itp.ac.cn}
\affiliation{College of Science, Xi'an University of Science and Technology, Xi'an 710054, People's Republic of China}
\begin{abstract}
In this work, a way starting from beta function is presented for obtaining well-defined coupling constant in UV and IR region.  In the approach presented here, obvious singularity is removed, and asymptotic behaviour is reserved fully and manifestly. Also it's shown that the freezed coupling constant is independent of dimensional parameters $\Lambda$ and not sensitive to higher-loop corrections and renormalization scheme adopted. A non-dimensional function of energy scale is introduced to play the role of ``fine tuning''.
\end{abstract}
\maketitle
\section{INTRODUCTION}
\label{INTRODUCTION}
The knowledge about behaviour of coupling constant in all range is of fundamental importance. RG or RGE is extensively used to obtain this
knowledge. However we must take care of constraint of expression obtained from RG or RGE. Whether our resultant expression for coupling constant is reliable depends on whether it is in accordance with constraint or not. It's almost always the case that our expression for coupling constant blows up or becomes enough large in some region or point where perturbation breaks down. Therefore, it's meaningful to find a method to remedy this. This meaningful way must own property that in safe region it does not change the behaviour of coupling constant effectively, in dangerous region it provides a well-defined expression to coupling constant.

As is well known, in 1950s Landau pointed out that in QED there was a scale at which our perturbation method failed. In QED this
scale is so high that we can ignore this from the point of view of effective field theory. However, in QCD the scale from which perturbation method fails lies in the quite physical IR region. The freezing of coupling constant provides a way out of this dilemma. In \cite{Banks:1981nn,Gardi:1998rf,Gardi:1998ch,Chishtie:1999tx,Ryttov:2010iz,Shrock:2013pya,Shrock:2013ca,Pica:2010xq,Ryttov:2016ner,Xue:2016dpl,Xue:2015wha,Xue:2016txt} the freezing of coupling constant in IR region is interpreted as a result of vanishing of beta function (of course the exact freezing value depends on the explicit form of beta function). Also, the freezing of coupling constant as a result of vanishing of beta function in UV region has been studied too \cite{Shrock:2014zca,Shrock:2013cca,Xue:2016txt}, for example, recently this has been studied to six-loop order for $\lambda\phi^4$ theory in \cite{Shrock:2016hqn}.

In \cite{Shirkov:1997wi} a way to extract behaviour of QCD coupling constant in IR region has been found and used, and it was concluded that coupling constant in IR region was freezed at a finite value (especially $\alpha_s(0)$ is universal) by using ``analytization procedure''.
The ``analytization procedure'' elaborated in \cite{Redmond:1958juf,Redmond:1958pe,trio} and used in \cite{Shirkov:1997wi} includes three steps:\\
(I) Finding an explicit expression for $\alpha_s(\mu^2)$ in the
Euclidean region;\\
(II)Performing analytical continuation into the Minkowski region. Extracting its imaginary part for defining the spectral
density---by $\rho_{RG} (\sigma ,\alpha)= \text{Im}\alpha(-\sigma-i\epsilon ,\alpha)$;\\
(III) Using $\rho_{RG}$ to define a new obtained ``analytically-improved''
running coupling constant in the Euclidean region.

The spirt of this approach is to use analyticity method to subtract singularity or Landau pole. However it's not manifest that the well-behaviour property and meaning of coupling constant are maintained by this way. And expression for coupling constant must be obtained firstly by direct calculation or solving beta function, but both of them are difficult in some cases.

The $\beta$-function for coupling constant may be obtained easily using some elegant procedure. For example, in the background field method, to calculate beta function we only need to  calculate the background field renormalization constant $Z_{A}$ \cite{Abbott:1980hw,Abbott:1983zw,Abbott:1981ke,Weinberg:1996kr,Herzog:2017ohr}. Recently, methods have been proposed to extract an all-orders $\beta$-function \cite{Pica:2010mt,Ryttov:2007cx}. The significant property that the explicit form of $\beta$-function (in mass-independent renormalization  schemes) only depends on term $1/{\epsilon}$ brings simplification further.

In this work, we start from $\beta$-function to obtain an equation about coupling constant and energy scale taking form such as
\begin{equation}
\frac{1}{(a_0\ln \alpha(\mu)+\sum_{i\neq0} a_i\alpha(\mu)^i)}=\frac{2}{\ln\frac{\mu^2}{\Lambda^2}}.\nonumber
\end{equation}
Using Cauchy theorem we can remove singularity of this expression; and then new coupling constant which is free of singularity (Landau pole or ghost pole) and has well-defined meaning in all region can be obtained. Also the main property of coupling constant can be kept fully and manifestly, such as asymptotic behaviour, and freezing of coupling constant is manifest.

This paper is organised as follows. We begin in Sec. II with a brief presentation and discussion of Cauchy theorem, which will be extensively used in this work. In Sec. III we use this to remove singularity to get a improvement of coupling constant. In Sec. IV, fixed point of coupling constant is discussed. Sec. V, contains a further analysis about fixed point (or freezing coupling constant), also in this section a non-dimensional function of energy scale is introduced to play the role of ``fine tuning''. Discussion and conclusion are presented in Sec. VI.
\section{CAUCHY THEOREM}
\label{CAUCHY THEOREM}
It's known that if a function is analytic in a connected region surrounded by a contour, we can use Cauchy theorem
\begin{eqnarray}
\oint f(z)\ud z&=&2\pi i\sum_i Resf(z_i)\\
f(z)&=&\frac{1}{2\pi i}\oint\frac{f(x)}{x-z}\ud z.
\end{eqnarray}
Also, in some cases Cauchy theorem can be used to extract the analytic part of a given function (some simple cases are presented in Appendix).

In the expression of coupling constant logarithm terms often appear as a result of the form of beta function, which often indicates the break down of perturbation theory. In
this work this problem is circumvented by inverting logarithm function and some analyticity procedure. So now we concentrate on this. A function with form
$\frac{1}{\ln\frac{x}{a}}$ has a isolate singularity point $a$ and a branch cut on negative real axis (assuming $a>0$). Note that we can
subtract this singularity point using the Cauchy theorem. Since we just want to get a new function defined in region $x>0$, using Cauchy theorem (the contour we choose is shown in Figure.1), we get
\begin{equation}
\tilde{f}(x)=\frac{1}{\ln\frac{x}{a}}+\frac{a}{a-x}
\end{equation}
which is obviously free of singularity pole $a$.
\begin{figure}
\includegraphics[scale=0.5]{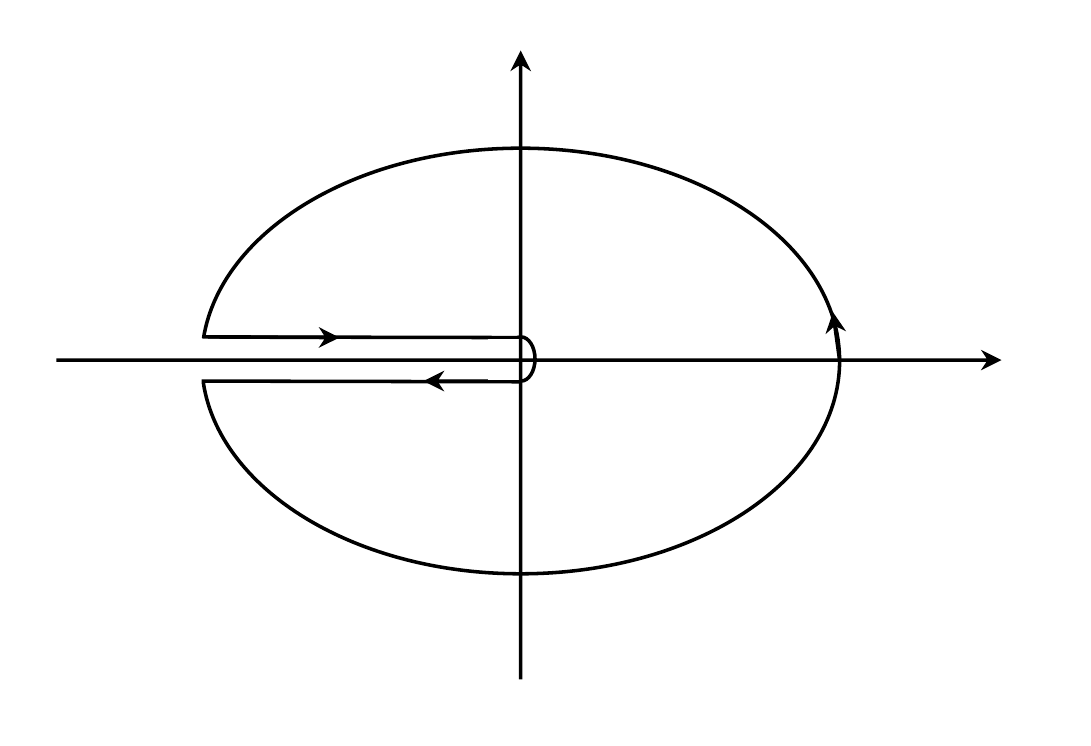}
\caption{Cauchy Theorem}
\end{figure}

\section{IMPROVEMENT OF COUPLING CONSTANT}
\label{IMPROVEMENT OF COUPLING CONSTANT}
In this section a method, which in QCD fully keeps behaviour of coupling constant in UV region and gives a well-defined expression to coupling constant in IR region, which in QED fully keeps behaviour of coupling constant in IR region and gives a well-defined expression to coupling constant in UV region, is presented.

The start point of this method is beta function which generally takes the form
\begin{equation}
\beta(\alpha)=\frac{\ud \alpha}{\ud \ln\mu}=\beta_0\alpha^{n_0}+\beta_1\alpha^{n_1}+\beta_2\alpha^{n_2}+\ldots
\end{equation}
which can be rewritten as
\begin{equation}
\beta(\alpha)=\frac{\ud \alpha}{\ud \ln\mu}=\frac{\beta_0\alpha^{n_0}}{1+\beta'_1\alpha^{n'_1}+\beta'_2\alpha^{n'_2}+\ldots}
\end{equation}
These new coefficient sets in Eq. (5) can be derived from matching with Eq. (4), and solution of Eq. (5) usually takes the form
\begin{equation}
a_0\ln \alpha(\mu)+\sum_{i\neq0} a_i\alpha(\mu)^i=\ln \frac{\mu}{\Lambda}
\end{equation}
$\Lambda$ is a integral constant. Inverting Eq. (6), we get
\begin{equation}
\frac{1}{a_0\ln \alpha(\mu)+\sum_{i\neq0} a_i\alpha(\mu)^i}=\frac{2}{\ln \frac{\mu^2}{\Lambda^2}}
\end{equation}
when Cauchy theorem described above is used, the r.h.s. of Eq. (7) change into
\begin{equation}
\tilde{f}(\mu^2)=2(\frac{1}{\ln\frac{\mu^2}{\Lambda^2}}+\frac{\Lambda^2}{\Lambda^2-\mu^2})
\end{equation}
which is obviously free of singularity and leads to our final equation
\begin{equation}
\frac{1}{a_0\ln \alpha(\mu)+\sum_{i\neq0} a_i\alpha(\mu)^i}=2(\frac{1}{\ln\frac{\mu^2}{\Lambda^2}}+\frac{\Lambda^2}{\Lambda^2-\mu^2})
\end{equation}
(here and below, for compactness and simplicity, we use $\alpha(\mu)$ to denote our new improved coupling constant) through which a exact solution for running coupling constant can be obtained (in some simple cases), if not the case, numerical solution can be get at least.

Note that $\tilde{f}(\mu^2)$ which ranges from $0$ to $2$ in Eq. (8) is a positive monotonous decreasing function of $\mu^2$. In this work Eq. (9) is used in case such
as QCD (ultraviolet-free theory) which has asymptotic behaviour in high energy scale. Using Eq. (9) we can define behaviour of coupling constant in IR region well. However
it's not suit for theory such as QED (infrared well-defined theory), which has asymptotic behaviour in IR region, and in which we want to define behaviour of coupling constant in UV
region meaningfully. The problem lies in the inversion of $\tilde{f}(\mu^2)$ which leads to infinity in UV region---$a_0\ln \alpha(\infty)+\sum_{i\neq0} a_i\alpha(\infty)^i$ $\rightarrow$ $\infty$, which may cause non-analyticity.

Another similar expression as Eq. (8) which is suit for infrared well-defined theory---QED---can be obtained as
\begin{equation}
\tilde{f_1}(\mu^2)=-2[\frac{1}{\ln\frac{\Lambda^2}{\mu^2}}+\frac{\mu^2}{\mu^2-\Lambda^2}]
\end{equation}
through which a similar equation as Eq. (9) is obtained as
\begin{equation}
\frac{1}{a_0\ln \alpha(\mu)+\sum_{i\neq0} a_i\alpha(\mu)^i}=-2[\frac{1}{\ln\frac{\Lambda^2}{\mu^2}}+\frac{\mu^2}{\mu^2-\Lambda^2}]
\end{equation}
$\tilde{f_1}(\mu^2)$ which ranges from $-2$ to $0$ in Eq. (10) obviously is a negative monotonous decreasing function of $\mu^2$. In this work Eq. (11) is suit for defining behaviour of coupling constant in UV region meaningfully.

$\Lambda^2/(\Lambda^2-\mu^2)$ in Eq. (8) is small compared with $1/{\ln\frac{\mu^2}{\Lambda^2}}$ in UV region. Therefore our way keeps
property of coupling constant in UV region well, and gives well-defined meaning to coupling constant in IR region.

${\mu^2}/(\mu^2-\Lambda^2)$ in Eq. (10) is small in IR region compared with $1/\ln\frac{\Lambda^2}{\mu^2}$. Therefore, our way keeps
property of coupling constant in IR region well, and gives well-defined meaning to coupling constant in UV region.

To illustrate our approach, let's use it to one-loop case.

One-loop order beta function for QCD:
\begin{equation}
\frac{\ud g(\mu)}{\ud\ln\mu}=-\beta_0\frac{g^3(\mu)}{16\pi^2}
\end{equation}
form which we get
\begin{equation}
\alpha_s(\mu)=\frac{4\pi}{\beta_0}[\frac{1}{\ln\frac{\mu^2}{\Lambda^2}}+\frac{\Lambda^2}{\Lambda^2-\mu^2}]
\end{equation}
which obviously has well-defined behaviour in IR region, and freezes at a limit value $\alpha_s(0)=\frac{4\pi}{\beta_0}$ (the same result has been given in \cite{Shirkov:1997wi})

One loop order beta function for QED:
\begin{equation}
 \frac{\ud e(\mu)}{\ud\ln\mu}=\frac{e^3(\mu)}{12\pi^2}
\end{equation}
from which we get
\begin{equation}
\frac{e^2(\mu)}{4\pi}=3\pi[\frac{1}{\ln\frac{\Lambda^2}{\mu^2}}+\frac{\mu^2}{\mu^2-\Lambda^2}]
\end{equation}
which obviously has well-defined behaviour in UV region, and freezes at a limit value $\alpha_s(\infty)=3\pi$ (the same result has been given in \cite{Shirkov:1997wi,trio})

To higher-loop order it's not such a easy work to extract a explicit expression for coupling constant in terms of $\mu^2$ explicitly. If the original beta function can be solved, we can add a perturbation term to it's solution, if not, we can at least get a numerical solution.

In the approach of this work we first subtract singularity then get coupling constant, however in \cite{Shirkov:1997wi} the coupling constant is obtained firstly and then analytical procedure to subtract singularity is used. The coincidence---in QCD or QED, the leading order result obtained in this work is exactly the same as that given in \cite{Shirkov:1997wi}---can be traced back to the absence of difference between this two way in simple case (only one coupling constant term appears in l.h.s. of Eq. (12) or Eq. (14)). For higher-loop order the difference between this two way emerges.

\section{FIXED POINT OF COUPLING CONSTANT}
\label{FIXED POINT OF COUPLING CONSTANT}
In \cite{Shirkov:1997wi,trio}, it was shown that the freezing coupling constant is independent of experimental estimates, which can be verified evidently in this work to any given order (even non-perturbative). The limiting value of $2(\frac{1}{\ln\frac{\mu^2}{\Lambda^2}}+\frac{\Lambda^2}{\Lambda^2-\mu^2})$ in Eq. (9) in IR region is independent of $\Lambda$ and renormalization scheme (this limiting value is $2$), and the limiting value of $-2[\frac{1}{\ln\frac{\Lambda^2}{\mu^2}}+\frac{\mu^2}{\mu^2-\Lambda^2}]$ in Eq. (11) in UV region is independent of $\Lambda$ and renormalization scheme (this limiting value is $-2$). That's to say, the freezing coupling constant is independent of experimental estimates.

Also, in this work the existence of fixed point (freezing coupling constant), we think, is independent of renormalization scheme under some general condition (in our opinion, the existence of fixed point is guaranteed by a obvious reason---the existence of this two limiting value given above in IR (or UV) region implies the existence of corresponding fixed point, some details about fixed point will be discussed further in next section).

\section{FINE TUNING}
\label{FINE TUNING}
Note that $\Lambda^2/(\Lambda^2-\mu^2)$ in Eq. (8) is comparable with $1/{\ln\frac{\mu^2}{\Lambda^2}}$ when $\mu^2$ is in the vicinity of $\Lambda^2$ (for example when $\mu^2=\lambda\Lambda^2$, $10<\lambda<30$). Hence this added terms $\Lambda^2/(\Lambda^2-\mu^2)$ change the behaviour of coupling constant significantly, which may be not wanted (in some cases).

Note that a term $t(\mu^2)$ added to both side of Eq. (6) can play the role of ``fine tuning''. Effect of this manipulation in l.h.s. of Eq. (6) is
amount to change $\Lambda$ to $\Lambda \exp(-t(\mu^2))$. Defining $t'(\mu^2)$ as $\frac{\ud t(\mu^2)}{\ud\mu^2}$, using the same approach described above, we get
\begin{multline}
\frac{1}{a_0\ln\alpha(\mu)+\sum_{i\neq0} a_i\alpha(\mu)^i+t(\mu^2)}=\\
2[\frac{1}{\ln\frac{\mu^2}{\Lambda^2\exp(-2t(\mu^2))}}
+\sum_{i=1}^j\frac{\mu_i^2}{(\mu_i^2-\mu^2)(1+2\mu_i^2t'(\mu_i^2))}]
\end{multline}
where $\mu_i^2$ is determined by condition:
\begin{equation}
\mu_i^2=\Lambda^2\exp(-2t(\mu_i^2))
\end{equation}
Here we simplemindedly use Cauchy theorem, but we should prove whether the singularity is removed and whether the result is meaningful. Fortunately, it can be proved that singularity is removed under some general condition and the result is meaningful(for detail proof see Appendix).

In case of QCD, for not spoiling of main property of coupling constant in UV region, it's demanded that in UV region the second terms in bracket of Eq. (16) is sufficiently small compared with the first terms in bracket. Now come to the first dilemma proposed at the beginning of this section. The smallness of the second terms in bracket of Eq. (16), provides a way out of this dilemma. Under this requirement, a fast oscillating function $t(\mu^2)$, which leads to $t'(\mu_i^2)$ being sufficient large, is an obvious choice. Combining this two requirement, $t(\mu^2)$ = ${\sin(k\mu^m)}/{\mu^n}$ ($k$ is a large number ,$m,n$ can be chosen to be integer number) might be a good choice.

A similar expression as Eq. (16) which is suit for infrared well-defined theory is obtained as
\begin{multline}
\frac{1}{a_0\ln\alpha(\mu)+\sum_{i\neq0}a_i\alpha(\mu)^i+t(\mu^2)}=\\
-2[\frac{1}{\ln\frac{\Lambda^2}{\mu^2\exp(2t(\mu^2))}}
+\sum_i^j\frac{\mu^2}{(\mu^2-\mu_i^2)(1+2\mu_i^2t'(\mu_i^2))}]
\end{multline}
where $\mu_i^2$ is determined by condition:
\begin{equation}
 \mu_i^2=\frac{\Lambda^2}{\exp(2t(\mu_i^2))}
\end{equation}
In case of QED, for not spoiling of main property of coupling constant in IR region, it's demanded that in IR region the second terms in bracket of Eq. (18) is sufficiently small compared with the first terms in bracket.

As is shown in previous section, the leading order results (for QCD $\alpha_s(0)=\frac{4\pi}{\beta_0}$, and for QED $\alpha_s(\infty)=3\pi$) given above are not small, which may indicate the break down of using beta function to extract fixed point (in some cases, for example in phenomenology QCD the freezing coupling constant is expected to range from $0.4$ to $1$ \cite{Godfrey:1985xj,Mattingly:1993ej,Mattingly:1992ud,Dokshitzer:1995zt,Dokshitzer:1995qm,Aguilar:2001zy})---the second dilemma.

As is well known, the existence (nature) of fixed point and the first two coefficients of expansion for $\beta$ are independent of renormalization scheme \cite{gross,Pokorski:1987ed} (all others are formally scheme-dependent). In this work the first two coefficients of set---$\left[a_{-n}, \ldots, a_{-1}, a_0, a_1, \ldots\right]$---in Eq. (9) or Eq. (11) are scheme-independent.

Now come to the second dilemma proposed above. A term $t(\mu^2)$ being added to both side of Eq. (6) can change the exact value of freezing of coupling constant. The existence of (physically) fixed point (freezing coupling constant) only depends on whether there is reasonable $\alpha$ satisfying the condition
\begin{equation}
 y(\alpha)=a_0\ln \alpha(\mu)+\sum_{i\neq0} a_i\alpha(\mu)^i+t(\mu^2)=\pm\frac{1}{2}
\end{equation}
Adjusting $t(\mu^2)$ (for example, $t(\mu^2)$ can be set as a constant function $t(\mu^2)\equiv t_0$), we can get a $\alpha$ being not enough large ($\alpha<1$), which satisfies the equation $a_0\ln \alpha+\sum_{i\neq0} a_i\alpha^i+t(\mu^2)=\pm\frac{1}{2}$. For $y(\alpha)$, the dominant contribution comes from leading terms (the added term $t(\mu)$ is introduced to realize fine tuning, thus it may be the dominant part  ) in expression of $y(\alpha)$ (these terms are arranged under the order---$\left[\alpha^{-n}, \ldots, \alpha^{-1}, \ln\alpha, \alpha, \ldots\right]$). Therefore, if in one scheme we get a fixed point $\alpha_1$ being not enough large, in another new scheme we can get a new fixed point which differ from it very little. The reason is that the first two coefficients (the coefficients of dominant terms) are still the same, so we can adjust fixed point a little to compensate the effect of changing coefficient set (changing renormalization scheme) and higher-loop corrections, in other words, in our approach the existence and exact freezing value of fixed point are not sensitive to renormalization scheme and higher-loop corrections---which guarantees the validity and reliability of our approach.

To summary, we can adjust the form of added term to fit with data, the knowledge of this added terms may brings deep understanding.

\section{DISCUSSION AND CONCLUSION}
In this paper, we consider elimination of singularity (Landau singularity or ghost pole) from expression such as $1/{\ln \frac{\mu^2}{\Lambda^2}}$, that's to say we just consider the case---r.h.s of Eq. (6) being treated as a explicit function of $\mu^2$. L.H.S. of Eq. (6) as an explicit function of $\alpha(\mu^2)$ but a implicit function of $\mu^2$ may bring other singularity. In our previous work (still unpublished in our writing of this paper), the effect of singularity (if exist) arising from l.h.s. of Eq. (6) being treated as implicit function of $\mu^2$ is removed effectively through dispersion relation. Hence in this work we boldly and simplemindedly think that the elimination of singularity from $1/{\ln \frac{\mu^2}{\Lambda^2}}$ is valid and sufficient. By just subtracting the singularity in $1/{\ln\frac{\mu^2}{\Lambda^2}}$, the singularity of running coupling constant is subtracted effectively.

The hypothesis of the freezing of coupling constant in IR region has been widely used and studied in QCD phenomenology \cite{Badalian:2001by,Eichten:1974af,Eichten:1979ms,Richardson:1978bt,Aguilar:2004td}. In this work, we construct a method to obtain the knowledge of coupling constant in UV and IR region, which can be used to a sorts of quantum field theory---sucn as QCD and QED. Through this method, we can keep property of coupling constant in ``safe'' region fully (for QCD the ``safe'' region is UV region, for QED IR region), while in ``dangerous'' region (for QCD the ``dangerous'' region is IR region, for QED UV region) we can ``regulate" coupling constant to give a well-defined meaning to it. Also by this method the meaning of coulpling constant is reserved manifestly, and freezing of coupling constant is manifest. In this work the property that the freezing coupling constant is independent of experimental estimates is proved. Here an adjustable parameter function of energy scale is introduced to realize ``fine tuning''. The explicit form of added terms determined by matching with experiment data may reveal deep understanding about running coupling constant.

From the indication of phenomenology, it's believed that the transition from high energy scale to low energy scale and transition form perturbative region to none-perturbative region are smooth \cite{Aguilar:2004td}. Our method for getting well-defined behavior of coupling constant in all energy range in some sense offers some reason to  support this belief. In \cite{Nishijima:1996ji} it's shown that color confinement is an inevitable consequence of BRS invariance and asymptotic freedom inherent in QCD---here the analyticity is the key factor relating UV and IR region. Analyticity reserved manifestly in this work is essential to physics, through which UV region and IR region can be related---information of UV region (or IR region) can be extracted from that of IR (or UV)region.

\section*{Acknowledgements}
We thank Prof Y.Q Chen for helpful discussions and suggestions on the manuscript. The work of Z.Y. Zheng is supported by The National Nature Science Foundation of China under Grant No. 11275242. The work of G. L. Zhou is supported by The National Nature Science Foundation of China under Grant No. 11647022.

\appendix
\section{}
Here, some examples illustrating the use of Cauchy theorem in this work are presented, and the validity of using Cauchy theorem in Eq. (16) or Eq. (18) is verified.

Some simple examples illustrating the use of Cauchy theorem to subtract singularity\\
(1): $f(x)=\frac{1}{x-a}$, the singularity term is itself so if we subtract this singular term we get $0$, which can be easily verified using Eq. (2)\\
(2): $f(x)=\frac{1}{(x-a)(x-b)}$ the singularity terms are $\frac{1}{x-a}$ and $\frac{1}{x-b}$. Using Cauchy theorem we obtain a new function $\tilde{f}(x)$ taking form
\begin{eqnarray}
\tilde{f}(x)&=&\frac{1}{(x-a)(x-b)}-\frac{1}{(b-a)(x-b)}\nonumber \\
&-&\frac{1}{(x-a)(a-b)}
\end{eqnarray}
which is free of singularity obviously.

Now, we consider a general function useful in this work which takes form
\begin{equation}
F(x)=\frac{1}{\ln\frac{x}{g(x)}}
\end{equation}
here, $g(x)$ is a real and positive function of $x$ ($x>0$) and can be written as $g(x)=e^{h(x)}$.

Using Cauchy theorem
\begin{equation}
\tilde{F}(z)=\frac{1}{2\pi i}\oint\frac{1}{x-z}\frac{1}{\ln\frac{x}{g(x)}}
\end{equation}
and procedure described in this work, we got
\begin{equation}
\tilde{F}(x)=\frac{1}{\ln\frac{x}{g(x)}}+\sum_i\frac{x_ig(x_i)}{(x_i-x)(g(x_i)-x_ig'(x_i))}
\end{equation}
where, the first term in r.h.s. of Eq. (A4) is the contribution of reside of integral function at pole $x$, the second term is that of reside of integral function at pole $x_i$ which is determined by condition
\begin{equation}
 x_i=g(x_i)
\end{equation}

In the near vicinity of $x_i$ we define
 \begin{eqnarray}
 g(x)&=&e^{h(x)}\\
 h(x)&=&\ln x+k(x)
 \end{eqnarray}
 using this definition $\tilde{F}(x)$ changes into
 \begin{equation}
 \tilde{F}(x)=-[\frac{1}{k(x)}-\sum_i\frac{1}{(x-x_i)k'(x_i)}]
 \end{equation}
 in the near vicinity of $x_i$, we have
 \begin{eqnarray}
 k(x_i)&=&0\\
 k(x)-k(x_i)&=&k'(x_i)(x-x_i)+\nonumber \\
 (x-x_i)^2 k''(x_i)/2&+& (x-x_i)^3 k'''(x_i)/6\ldots
 \end{eqnarray}
Thus r.h.s. of Eq. (A8) can be rewritten as
 \begin{eqnarray}
 &-&[\frac{1}{k'(x_i)(x-x_i)+\frac{k''(x_i)}{2}(x-x_i)^2\ldots}-\frac{1}{k'(x_i)(x-x_i)}]\nonumber \\
 &+&\sum_{k\neq i}\frac{x_kg(x_k)}{(x_k-x)(g(x_k)-x_ig'(x_k))}
 \end{eqnarray}
which can be simplified into
\begin{eqnarray}
\frac{k''(x_i)}{2k'^2(x_i)}&+&\sum_{k\neq i}\frac{x_kg(x_k)}{(x_k-x)(g(x_k)-x_kg'(x_k))}\nonumber \\
&+&\mathcal{O}(x-x_i)
\end{eqnarray}
which is free of singularity pole $x_i$ obviously and has a well-defined meaning (here, we have assumed that $k'(x_i)\neq0$).

Thus we finish our proof.
 

\begin{thebibliography}{}
\bibitem{Banks:1981nn}
  T.~Banks and A.~Zaks,
  Nucl.\ Phys.\ B {\bf 196}, 189 (1982).
\bibitem{Gardi:1998rf}
  E.~Gardi and M.~Karliner,
  Nucl.\ Phys.\ B {\bf 529}, 383 (1998).
\bibitem{Gardi:1998ch}
  E.~Gardi and G.~Grunberg,
  JHEP {\bf 9903}, 024 (1999).
\bibitem{Chishtie:1999tx}
  F.~A.~Chishtie, V.~Elias, V.~A.~Miransky and T.~G.~Steele,
  Prog.\ Theor.\ Phys.\  {\bf 104}, 603 (2000).
\bibitem{Ryttov:2010iz}
  T.~A.~Ryttov and R.~Shrock,
  Phys.\ Rev.\ D {\bf 83}, 056011 (2011).
\bibitem{Shrock:2013pya}
  R.~Shrock,
  Phys.\ Rev.\ D {\bf 87}, 116007 (2013).
\bibitem{Shrock:2013ca}
  R.~Shrock,
  Phys.\ Rev.\ D {\bf 87}, no. 10, 105005 (2013).
\bibitem{Pica:2010xq}
  C.~Pica and F.~Sannino,
  Phys.\ Rev.\ D {\bf 83}, 035013 (2011).
  \bibitem{Ryttov:2016ner}
  T.~A.~Ryttov and R.~Shrock,
  Phys.\ Rev.\ D {\bf 94}, no. 10, 105015 (2016).
\bibitem{Xue:2016dpl}
  S.~S.~Xue,
  JHEP {\bf 1611}, 072 (2016).
\bibitem{Xue:2015wha}
  S.~S.~Xue,
  Phys.\ Rev.\ D {\bf 93}, no. 7, 073001 (2016).
\bibitem{Xue:2016txt}
  S.~S.~Xue,
  arXiv:1601.06845 [hep-ph].
\bibitem{Shrock:2014zca}
  R.~Shrock,
  Phys.\ Rev.\ D {\bf 90}, no. 6, 065023 (2014).
\bibitem{Shrock:2013cca}
  R.~Shrock,
  Phys.\ Rev.\ D {\bf 89}, no. 4, 045019 (2014).
\bibitem{Shrock:2016hqn}
  R.~Shrock,
  Phys.\ Rev.\ D {\bf 94}, no. 12, 125026 (2016).
\bibitem{Shirkov:1997wi}
  D.~V.~Shirkov and I.~L.~Solovtsov,
  Phys.\ Rev.\ Lett.\  {\bf 79}, 1209 (1997).
\bibitem{Redmond:1958juf}
  P.~J.~Redmond,
  Phys.\ Rev.\  {\bf 112}, no. 4, 1404 (1958).
\bibitem{Redmond:1958pe}
  P.~J.~Redmond and J.~L.~Uretsky,
  Phys.\ Rev.\ Lett.\  {\bf 1}, 147 (1958).
  \bibitem{trio}  N.N. Bogoliubov, A.A. Logunov and D.V. Shirkov,  Sov. Phys.
JETP {\bf 37}(10) 574 (1960).
\bibitem{Abbott:1980hw}
  L.~F.~Abbott,
  Nucl.\ Phys.\ B {\bf 185}, 189 (1981).
\bibitem{Abbott:1983zw}
  L.~F.~Abbott, M.~T.~Grisaru and R.~K.~Schaefer,
  Nucl.\ Phys.\ B {\bf 229}, 372 (1983).
\bibitem{Abbott:1981ke}
  L.~F.~Abbott,
  Acta Phys.\ Polon.\ B {\bf 13}, 33 (1982).
\bibitem{Weinberg:1996kr}
  S.~Weinberg,
  The quantum theory of fields. Vol. 2: Modern applications.
\bibitem{Herzog:2017ohr}
  F.~Herzog, B.~Ruijl, T.~Ueda, J.~A.~M.~Vermaseren and A.~Vogt,
  JHEP {\bf 1702}, 090 (2017).
\bibitem{Ryttov:2007cx}
  T.~A.~Ryttov and F.~Sannino,
  Phys.\ Rev.\ D {\bf 78}, 065001 (2008).
\bibitem{Pica:2010mt}
  C.~Pica and F.~Sannino,
  Phys.\ Rev.\ D {\bf 83}, 116001 (2011).
\bibitem{Godfrey:1985xj}
  S.~Godfrey and N.~Isgur,
  Phys.\ Rev.\ D {\bf 32}, 189 (1985).
\bibitem{Mattingly:1993ej}
  A.~C.~Mattingly and P.~M.~Stevenson,
  Phys.\ Rev.\ D {\bf 49}, 437 (1994).
\bibitem{Mattingly:1992ud}
  A.~C.~Mattingly and P.~M.~Stevenson,
  Phys.\ Rev.\ Lett.\  {\bf 69}, 1320 (1992).
\bibitem{Dokshitzer:1995zt}
  Y.~L.~Dokshitzer and B.~R.~Webber,
  Phys.\ Lett.\ B {\bf 352}, 451 (1995).
\bibitem{Dokshitzer:1995qm}
  Y.~L.~Dokshitzer, G.~Marchesini and B.~R.~Webber,
  Nucl.\ Phys.\ B {\bf 469}, 93 (1996).
\bibitem{Aguilar:2001zy}
  A.~C.~Aguilar, A.~Mihara and A.~A.~Natale,
  Phys.\ Rev.\ D {\bf 65}, 054011 (2002).
   \bibitem{gross}
  D.~J.~Gross,
  in Methods in Field Theory,Les Houches 1975, edited by R.Balian and J. Zinn-Justin (North Holland, Amsterdam, 1976), p.141.
\bibitem{Pokorski:1987ed}
  S.~Pokorski,
  Gauge Field Theories.
\bibitem{Badalian:2001by}
  A.~M.~Badalian and D.~S.~Kuzmenko,
  Phys.\ Rev.\ D {\bf 65}, 016004 (2001).
\bibitem{Eichten:1974af}
  E.~Eichten, K.~Gottfried, T.~Kinoshita, J.~B.~Kogut, K.~D.~Lane and T.~M.~Yan,
  Phys.\ Rev.\ Lett.\  {\bf 34}, 369 (1975)
  Erratum: [Phys.\ Rev.\ Lett.\  {\bf 36}, 1276 (1976)].
\bibitem{Eichten:1979ms}
  E.~Eichten, K.~Gottfried, T.~Kinoshita, K.~D.~Lane and T.~M.~Yan,
  Phys.\ Rev.\ D {\bf 21}, 203 (1980).
\bibitem{Richardson:1978bt}
  J.~L.~Richardson,
  Phys.\ Lett.\  {\bf 82B}, 272 (1979).
\bibitem{Aguilar:2004td}
A.~C.~Aguilar, A.~Mihara and A.~A.~Natale,
Int.\ J.\ Mod.\ Phys.\ A {\bf 19}, 249 (2004).
\bibitem{Nishijima:1996ji}
  K.~Nishijima,
  Czech.\ J.\ Phys.\  {\bf 46}, 1 (1996).
  \end{thebibliography}
\end{document}